\documentclass[useAMS,usenatbib]{mn2e}
\usepackage{graphicx}
\usepackage{amssymb}

\newcommand{\etal}{et al.~}

\def\chandra{{\sl Chandra}}

\def\gtrsim{\lower 2pt \hbox{$\, \buildrel {\scriptstyle >}\over
{\scriptstyle \sim}\,$}}
\def\lesssim{\lower 2pt \hbox{$\, \buildrel {\scriptstyle <}\over
{\scriptstyle \sim}\,$}}

\def\oviii{O~{\scriptsize VIII}}
\def\ovii{O~{\scriptsize VII}}

\def\ka{K{$\alpha$}}

\usepackage{fixltx2e}
\begin{document}

\title[]{Comments on "A huge reservoir of ionized gas around the Milky Way: accounting for the missing mass?" (2012 ApJL, 756, 8) and "The warm-hot gaseous halo of the Milky Way" (arXiv1211.3137)}
\author[]{Q. Daniel Wang$^{1}$\thanks{E-mail:wqd@astro.umass.edu} and Yangsen Yao$^{2,3}$\\
$^{1}$Department of Astronomy, University of Massachusetts, 
  Amherst, MA 01003, USA\\
$^{2}$Center for Astrophysics and Space Astronomy, Department of Astrophysical and Planetary Sciences, University of Colorado, Boulder, CO 80309\\
$^{3}$Eureka Scientific, Oakland, CA 94602}


\maketitle

\begin{abstract}
The two papers referred to in the title, claiming the detection of a large-scale massive hot gaseous halo around the Galaxy, have generated a lot of confusion and unwarranted excitement (including public news coverage). However, the papers are seriously flawed in many aspects, including problematic analysis and assumptions, as well as mis-reading and mis-interpreting earlier studies, which are inconsistent with the claim. Here we show examples
of such flaws.
\end{abstract}


\section{INTRODUCTION}
There are lines of evidence for a warm-hot gaseous halo around the Milky Way Galaxy (e.g., Wang 2010; Putman et al. 2012; 
and references therein). But how extended and massive this halo remains
a serious question to address. In a recent paper (Paper I hereafter), Gupta et al. (2012) have argued that the halo is extended over a region with a radius larger 
than 100 kpc and that the mass content of the gas with 
a temperature of $\sim 2 \times 10^6$ K is greater than  $10^{10} M_\odot$, comparable to the total baryonic
mass in the Galactic disk and accounting for the bulk of the missing baryon matter in the 
Galaxy. Their argument is based on an inference from a {\sl mean} \ovii\ column density obtained from X-ray absorption line measurements of \ovii\ \ka\ and K$\beta$ transitions, together with an {\sl average} emission measure (EM) estimated from previous studies of the soft X-ray background in 
{\sl different} parts of the sky. This inference critically 
assumes that both the line absorption and the background emission arise from the same plasma with a single temperature and a single metal abundance. 
As will be detailed in the following, both the measurement procedure and 
the simplistic inference of the work are very problematic. The results and conclusions 
are inconsistent with many previous studies (e.g., Wang et al. 2005; Yao \& Wang 2005, 
2007a; Bregman \& Lloyd-Davies 2007; Yao et al. 2009), some of which are 
based on more or less the same data sets.

Indeed, Mathur (2012; Paper II hereafter) very recently reviewed some of these 
inconsistencies, mostly by criticizing some of these
previous studies. Unfortunately, this review is highly biased 
and in fact contains numerous errors due to mis-reading and/or 
mis-interpreting the studies. 

These problems and errors together very much explain the inconsistencies.
In the following, we show examples of these, 
based on quick fact-checking and analysis of relevant issues.

\section{Examples of Problems and Errors}

\begin{figure}
\centering
\includegraphics[width=3.0in]{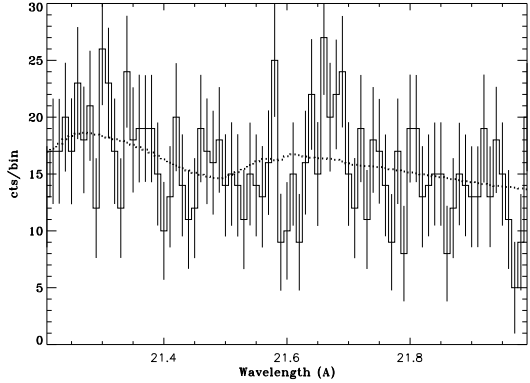}
\caption{An example spectrum of Ark564 around the \ovii\ K$\alpha$ line (21.6 \AA). The spectrum represents the combination of the MEG spectra 
(downloaded from TGCAT(http://tgcat.mit.edu/) extracted from all existing
\chandra/HETG observations (ObsID 863, 9898, 9899, and 10575) 
with a total effective exposure of 296~ks. The dotted curve represents
a fit of a power law (convolved with the instrument response) 
to the spectrum. 
}
\label{f:ark564}
\end{figure}

Let us first concentrate on Paper I. 
While the data analysis description of 
the paper is not sufficiently detailed for us to reproduce their results exactly, Fig.~1 of Paper I clearly indicates problems. The dispersion of the data points in each of the eight panels, showing {\sl normalized} flux
spectra of eight AGNs, is substantially smaller than the (presumably 1 $\sigma$) error bars. 
In comparison, Fig.~\ref{f:ark564} here shows an original spectrum of Ark 564, covering the same wavelength range as was presented in the paper. 
It is hard to believe that there is an absorption line of an equivalent width
(EW) of $ 12.0 \pm 1.9$, as claimed in Table 2 of the paper. We have learned that many of the apparent "peaks" and "valleys"  (except for the bin(s) at 21.6 \AA) 
in the original spectrum have been modeled as being intrinsic to the AGN and thus have disappeared in the normalized flux spectra (data/model) plot, which explains the low
dispersion among the data points. In addition, most of the claimed new detections of the ($z \sim 0$) absorption lines 
are apparently too narrow and/or too deep (reaching zero fluxes in 
the normalized spectra) to be 
consistent with the limited resolution of the instruments. Therefore, the 
very detections of the lines 
are questionable, let alone their quoted high significances.

It is also not appropriate to adopt the mean values of the X-ray absorption 
column densities and EMs to infer the physical scale of the halo,
because of the large variation of these measurements from one region to another 
(e.g., Table~2 in Paper I; Yoshino et al. 2009);  the EM (after 
the solar wind charge exchange contribution is subtracted), 
for example, varies by a factor of $\gtrsim 10$ and shows
a strong Galactic latitude dependence. The patchiness and latitude
dependence, as seen in both the X-ray emission and absorption  
(e.g., Yao \& Wang 2005; Yoshino et al. 2009), can hardly be
explained by the claimed large-scale, smooth, massive, hot gaseous halo.
The emission and absorption comparison should  be
made for adjacent fields of the AGN sight lines, as 
have been performed in existing studies (e.g., Yao et al. 2009; 
Hagihara et al. 2010, 2011; Sakai et al. 2012).

\begin{figure}
\centering
\includegraphics[width=3.0in]{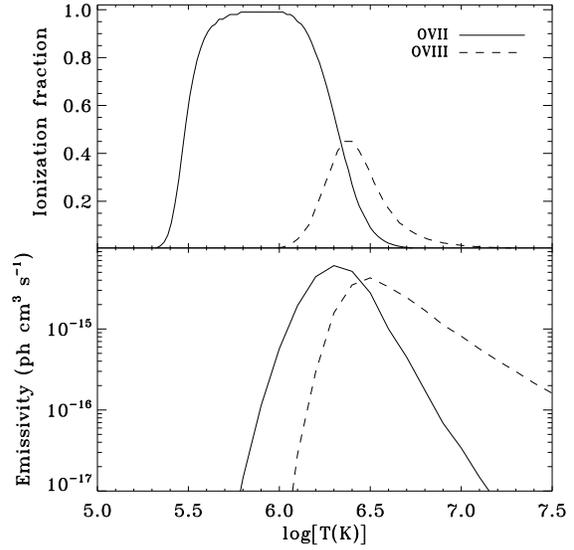}
\caption{Temperature dependence of the oxygen ionic fractions 
(upper panel; Gnat, O., \& Sternberg, A. 2007) and oxygen line emissivities for a CIE
plasma (lower panel; Foster et al. 2012). Notice the logarithmic scale of the
emissivity.}
\label{f:ionfr_emis}
\end{figure}

But more problematic is the use of the single temperature plasma assumption 
(implicitly made in Paper I)
to infer the scale of the halo. This assumption has been shown 
to be inconsistent with the combined absorption and 
emission data (e.g., Yao \& Wang 2007a, Yao et al. 2009; Hagihara et al. 2010, 2011; Sakai et al. 2012). The absorption lines tend to sample plasma 
over a lower temperature range than the emission.
The \ovii\ K$\alpha$ line emissivity peaks strongly at $\sim 10^{6.3}$~K  and drops by more than
an order of magnitude at $10^{6}$ K, for example, whereas the \ovii\ \ka\ absorption line
samples a broad lower temperature range (the ionic fraction of \ovii\ 
is $\gtrsim 0.5$ over
$10^{5.5}-10^{6.3}$ K; Fig.~\ref{f:ionfr_emis}). 
The characteristic temperature as obtained from the \ovii/\oviii\
absorption line ratio, for example, predicts a far too larger \ovii/\oviii\ emission line ratio than observed (e.g., Yao \& Wang 2007a). The existing measurements show that the temperatures inferred from the fits of multiple absorption lines are systematically lower than those from emission spectral analyses. Again, such comparisons need to be done for
the adjacent emission and absorption pairs. Ignoring the different 
temperature sensitivities of the emission and absorption measurements 
can naturally lead to an over-estimation of the effective halo scale by more than an order of magnitude --- an important effect that Paper I gives 
no discussion to.

Let us now turn to Paper II. Problems in this paper are numerous. We
choose to rebut a few key sentences quoted directly from the paper:

\begin{itemize}

\item Paper II states "we have also discussed all the assumptions and biases clearly".

As described above, the effect due to the single temperature assumption, which is also not discussed in the paper, is likely the most important factor affecting the scale estimation of the halo.

\item On the \ovii\ K$\alpha$ absorption line toward LMC X-3, which at a distance of only 50 kpc shows an \ovii\ \ka\ absorption consistent with those seen in 
AGNs (Wang et al. 2005; Bregman \& Lloyd-Davies 2007; Yao et al. 2009), Paper II states "the \ovii\ EW is not well constrained; it is consistent with zero at 2.7$\sigma$."

We have no idea as to how this 2.7$\sigma$ is derived. Table~1 of Wang et al. (2005) lists the EW of the \ovii\ K$\alpha$ line as 20(14, 26) m\AA, where 14-26 is the 90\% uncertainty interval. Therefore, the EW is $20_{-6}^{+6}$~m\AA\ (90\% error bars), not $20_{-26}^{+14}$~m\AA, as quoted in Paper II, which simply does not make any sense. With this mistake, the Paper I's subsequent argument on 
this issue is pointless. In fact, similar differential X-ray absorption line 
analysis has been carried out for other sight lines  through the Galactic halo (e.g., Yao \& Wang 
2005, 2007a,b) and through the halos of 
other galaxies (Yao et al. 2010). No evidence for large-scale, {\sl massive} gaseous halos is
revealed.

\item On the comparison between the Mrk 421 and 4U 1957 + 11 sightlines (Yao et al. 2008), Paper II  states "the halo contribution is at least $10 - 3 = 7 \times 10^{15} \rm~cm^{-2}$".

This quoted derivation without incorporating the large error bars is meaningless.  Furthermore, the used column density N(\ovii) values for the 4U source are already corrected for the Galactic latitude dependence, as described clearly in the note to Table 1
of Yao et al. (2008):  "The dependence of the column density on the Galactic latitude (sin b factor) has been corrected with respect to the Mrk 421 direction". Before the correction, the directly measured N(\ovii) is a factor of 5 higher. Thus, the meaning of the numbers 
is misread and the subsequent criticism is completely baseless in Paper II.

\item Paper II also states "their own analysis shows that this assumption is invalid; they find the b-parameter in the 4U line to be 155 km/s, while in the Mrk 421 line it is 64 km/s, therefore the two spectra should not be fit together."

Again quoting the best-fit numbers alone without including the error bars (let alone making a conclusion out of the comparison) is very misleading. Table 1 of Yao et al. (2008) lists the b values as 64 (48, 104) and 155 (70, 301) km/s for the sight lines toward Mrk 421 and 4U. Therefore, the b values are not well 
constrained for the individual sight lines; their 90\% intervals are well
overlapped. It is thus not unreasonable to test the disk model, 
by jointly fitting the b value for both sight lines. Of course, this is only a consistency check. The motivation for the model and other consistency checks are presented in several other papers (e.g., Yao \& Wang 2005, 2007a), which Paper II completely ignore.

\item Paper II further states "the new column toward 4U they find is more than the 90\% upper limit on this column from the 4U spectrum alone. This cannot be right and is most likely the result of a much lower b-value (= 70 km/s) in the joint fit than in the 4U spectrum alone"

The quoted 90\% error interval based on a single parameter estimate does not fully reflect the error range when the two parameters (the column density and the b value) are strongly correlated. So one should not be surprised to see the more tightly constrained fit with the joint b value gives a column density estimate that is just outside the 90\% internal from the fit to the single sight line.

\item Paper II states "All the papers to date have used an average emission measure for the halo; ideally we need emission measures close to the absorption sightlines."

As described above, such measurements have already been made with dedicated {\sl Suzaku} observations
for multiple sightlines, including PKS 2155-304 (Hagihara et al. 2010), 4U 1820-303 (Hagihara et al. 2011), Mkn 421 (Sakai et al. 2012),  and LMC X-3 (Yao et al. 2009). In these studies, the X-ray emission spectra extracted from adjacent fields are jointly analyzed with the X-ray grating absorption line data, all reaching the same conclusions: 1) the plasma is not isothermal and 2) its effective pathlength is small ($\lesssim 10$ kpc, partly depending on the plasma filling factor).  The conclusions were also reached earlier in the joint analysis of absorption line data with the {\sl ROSAT} all-sky survey intensities measured in the field around Mkn 421 (Yao \& Wang 2007a). The abstract of this paper clearly states ``We find that the observed absorption line strengths of \ovii\ and \oviii\ are inconsistent with the diffuse background emission-line ratio of the same ions, if the gas is assumed to be isothermal in a collisional ionization equilibrium state.'' We are astonished
that the author of Paper II has completely missed or ignored these references.
\end{itemize}

There are other problematic issues in Papers I and II, which 
ignore many important counter-arguments in existing work and are highly
biased in comparison with X-ray observations of nearby galaxies. These 
observations have established the presence of hot gas in and around many
nearby galactic disks and/or bulges (e.g., T\"ullmann et al. 2006; Wang 2010; Li \& Wang 2012). 
The presence of a similar hot gas component
on scales of a few kpc is also apparent in our Galaxy and must be accounted for
when the more distant and tenuous halo component is explored.


\section*{Acknowledgments}
We thank Dan McCammon, Chris Howk, and Mike Shull for soliciting our opinions on the work, which led to the comments presented above, Anjali
Gupta for helping to clarify key analysis steps in her work, and Dan McCammon, Todd Tripp 
and Mike Anderson for their comments on an early draft of this write-up.

\end{document}